\documentclass[conference]{IEEEtran}
\IEEEoverridecommandlockouts
\usepackage{graphicx}
\usepackage{amsmath}
\usepackage{listings}
\usepackage{breqn}
\usepackage{hyperref}
\usepackage{breqn}
\usepackage{graphicx}
\usepackage{caption}
\usepackage{subcaption}
\usepackage[margin=1cm]{caption}

\def\BibTeX{{\rm B\kern-.05em{\sc i\kern-.025em b}\kern-.08em
    T\kern-.1667em\lower.7ex\hbox{E}\kern-.125emX}}

\begin{document}

\title{Event Detection in Twitter: A Keyword Volume Approach}
\author{\IEEEauthorblockN{1\textsuperscript{st}Ahmad Hany Hossny}
\IEEEauthorblockA{\textit{School of Mathematical Sciences} \\
\textit{University of Adelaide}\\
Adelaide, Australia \\
ahmad.hossny@adelaide.edu.au}
\and
\IEEEauthorblockN{2\textsuperscript{nd} Lewis Mitchell}
\IEEEauthorblockA{\textit{School of Mathematical Sciences} \\
\textit{University of Adelaide}\\
\textit{Data to Decisions CRC stream lead}\\
lewis.mitchell@adelaide.edu.au}
}

\maketitle

\begin{abstract}
Event detection using social media streams needs a set of informative features with strong signals that need minimal preprocessing and are highly associated with events of interest. Identifying these informative features as keywords from Twitter is challenging, as people use informal language to express their thoughts and feelings. This informality includes acronyms, misspelled words, synonyms, transliteration and ambiguous terms. In this paper, we propose an efficient method to select the keywords frequently used in Twitter that are mostly associated with events of interest such as protests. The volume of these keywords is tracked in real time to identify the events of interest in a binary classification scheme. We use keywords within word-pairs to capture the context. The proposed method is to binarize vectors of daily counts for each word-pair by applying a spike detection temporal filter, then use the Jaccard metric to measure the similarity of the binary vector for each word-pair with the binary vector describing event occurrence. The top $n$ word-pairs are used as features to classify any day to be an event or non-event day. The selected features are tested using multiple classifiers such as Naive Bayes, SVM, Logistic Regression, KNN and decision trees. They all produced AUC ROC scores up to 0.91 and F1 scores up to 0.79. The experiment is performed using the English language in multiple cities such as Melbourne, Sydney and Brisbane as well as the Indonesian language in Jakarta. The two experiments, comprising different languages and locations, yielded similar results.
\end{abstract}


\section{Introduction}
Event detection is important for emergency services to react rapidly and minimize damage. For example, terrorist attacks, protests, or bushfires may require the presence of ambulances, firefighters, and police as soon as possible to save people. This research aims to detect events as soon as they occur and are reported via some Twitter user. The event detection process requires to know the keywords associated with each event and to assess the minimal count of each word to decide confidently that an event has occurred. In this research, we propose a novel method of spike matching to identify keywords, and use probabilistic classification to assess the probability of having an event given the volume of each word. 

Event detection and prediction from social networks have been studied frequently in recent years. Most of the predictive frameworks use textual content such as likes, shares, and retweets, as features. The text is used as features either by tracking the temporal patterns of keywords, clustering words into topics, or by evaluating sentiment scores and polarity. The main challenge in keyword-based models is to determine which words to use in the first place, especially as people use words in a non-standard way, particularly on Twitter.

In this research, we aim for detecting large events as soon as they happen with near-live sensitivity. For example, When spontaneous protests occur just after recent news such as increasing taxes or decreasing budget, we need to have indicators to raise the flag of a happening protest. Identifying these indicators requires to select a set of words that are mostly associated with the events of interest such as protests. We then track the volume of these words and evaluate the probability of an event occurring given the current volume of each of the tracked features. The main challenge is to find this set of features that allow such probabilistic classification.

\begin{figure}
\centering
\includegraphics[width=0.9\linewidth]{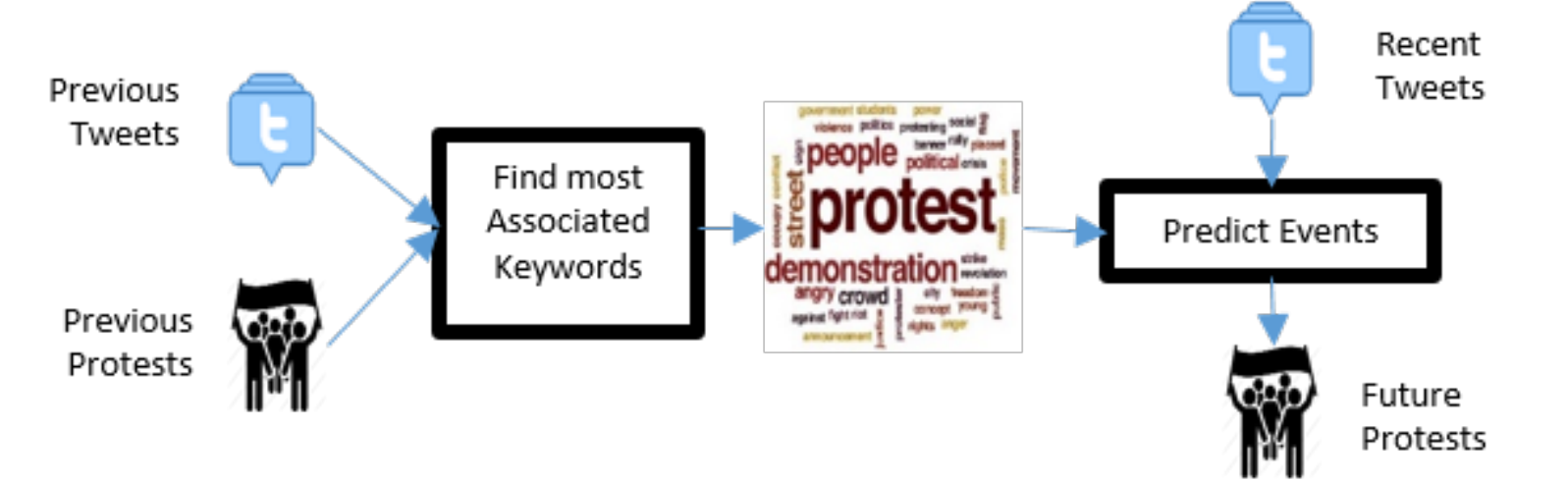}
\caption{The the proposed pipeline extracts the word-pairs matching events of interest, then use the extracted word-pairs as features to detect civil unrest events.}
\label{fig:pipeline}
\end{figure}

Using text as features in Twitter is challenging because of the informal nature of the tweets, the limited length of the tweet,  platform-specific language, and multilingual nature of Twitter \cite{fung2005parameter,Mathioudakis2010,Petrovic2010}. The main challenges for text analysis in Twitter are listed below:
\begin{enumerate}
    \item The usage of misspelled words, acronyms, an non-standard abbreviations make words and expression not understandable.   
    \item The transliteration of non-Latin languages such as Arabic using the Latin script distorts the feature signal where words with similar spelling have different meaning in different languages (e.g. the term ``boss'' in English means ``manager'', while in Arabic it means ``look'') .
    \item The limited length of the tweets makes sentiment analysis and topic modelling pretty  challenging.
    \item Semantic ambiguity: a single word can refer to many meanings (e.g. ``strike'' may refer to a lightning strike or a protest or a football strike)
\end{enumerate}

We approached the first and second challenges by using a Bayesian approach to learn which terms were associated with events, regardless of whether they are standard language, acronyms, or even a made-up word, so long as they match the events of interest. The third and fourth challenges are approached by using word-pairs, where we extract all the pairs of co-occurring words within each tweet. This allows us to recognize the context of the word ('Messi','strike' ) is different than ('labour','strike').  

\begin{figure}
\centering
\includegraphics[width=0.9\linewidth]{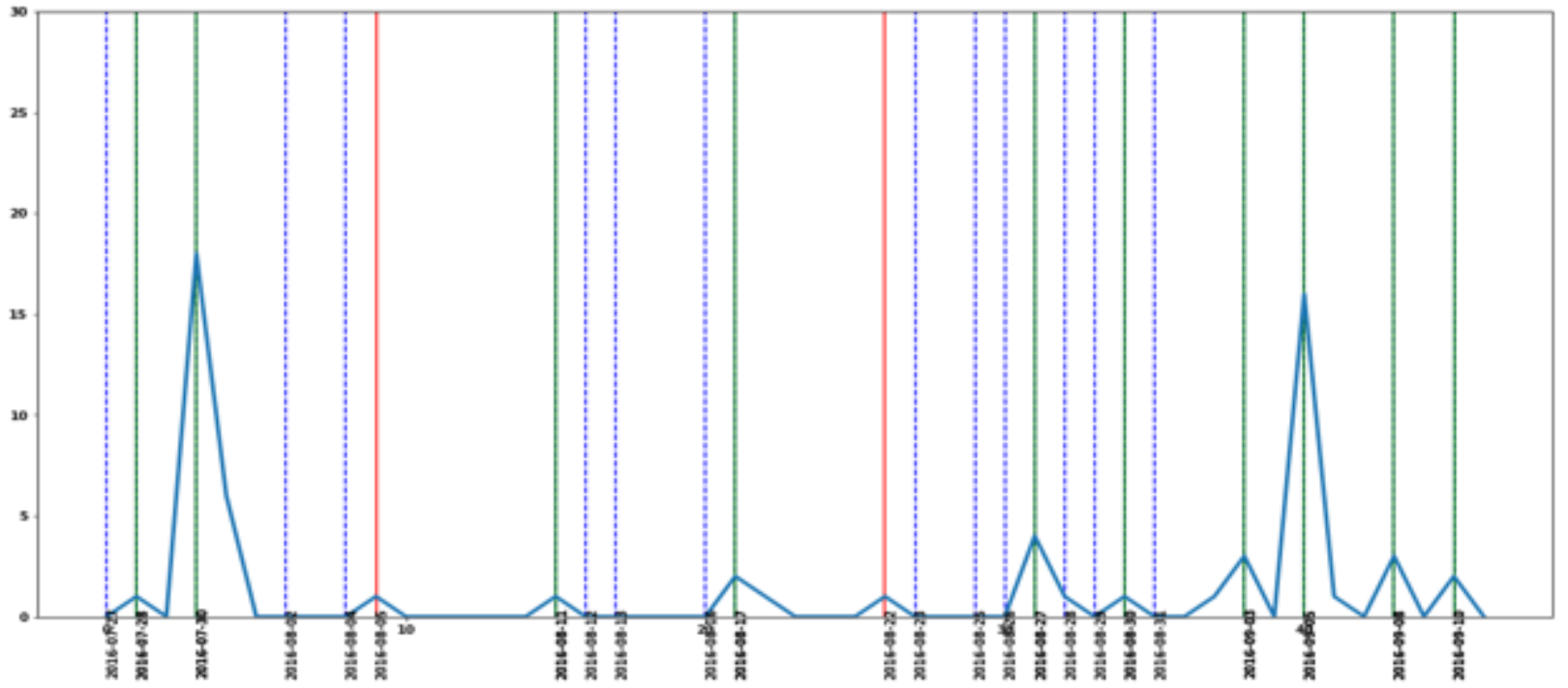}
\caption{The spikes in the time series signal for the word-pair (`Melbourne',`Ral') are matched with the event days that represented as dotted vertical lines. The green lines represent spikes matching events. The Blue lines represent events with no matching spikes and red lines represent spikes that did not match any event.}
\label{fig:SpikeMathcing}
\end{figure}

According to the distributional semantic hypothesis, event-related words are likely to be used on the day of an event more frequently than any normal day before or after the event. This will form a spike in the keyword count magnitude along the timeline as illustrated in Figure \ref{fig:SpikeMathcing}. To find the words most associated with events, we search for the words that achieve the highest number of spikes matching the days of events. We use the Jaccard similarity metric as it values the spikes matching events and penalizes spikes with no event and penalizes events without spikes. Separate words can be noisy due to the misuse of the term by people, especially in big data environments. So, we rather used the word-pairs as textual features in order to capture the context of the word. For example, this can differentiate between the multiple usages of the word ``strike'' within the contexts of ``lightning strike'', ``football strike'' and ``labour strike''

In this paper, we propose a method to find the best word-pairs to represent the events of interest. These word-pairs can be used for time series analysis to predict future events as indicated in Figure \ref{fig:pipeline}. They can also be used as seeds for topic modelling, or to find related posts and word-pairs using dynamic query expansion. The proposed framework uses a temporal filter to identify the spikes within the word-pair signal to binarize the word-pair time series vector \cite{lewicki1998review}. The binary vector of the word-pair is compared to the protest days vector using Jaccard similarity index \cite{niwattanakul2013using,bank2008calculating}, where the word-pairs with highest similarity scores are the most associated word-pairs with protest days. This feature selection method is built upon the assumption that people discuss an event on the day of that event more than on any day before or after the event. This implies that word-pairs related to the event will form a spike on this specific day. Some of the spiking word-pairs are related to the nature of the event itself, such as ``taxi protest'' or ``fair education''. These word-pairs will appear once or twice along the time frame. Meanwhile, more generic word-pairs such as ``human rights'' or ``labour strike'' will spike more frequently in the days of events regardless the protest nature.

To test our method, we developed two experiments using all the tweets in Melbourne and Sydney over a period of 640 days. The total number of tweets exceeded 4 million tweets per day, with a total word-pair count of 12 million different word-pairs per day, forming 6 billion word-pairs over the entire timeframe. The selected word-pairs from in each city are used as features to classify if there will be an event or not on a specific day in that city. We classified events from the extracted word-pairs using 9 classifiers including  Naive Bayes, Decision Trees, KNN, SVM, and logistic regression.

In Section 2, we describe the event detection methods. Section 3 states the known statistical methods used for data association and feature selection. Section 4 describes the proposed feature selection method. Section 5 describes model training and prediction. Section 6 describes the experiment design, the data and the results. Section 7 summarizes the paper, discuss the research conclusion and explains future work.

\section{Event Detection Methods}

Analyzing social networks for event detection is approached from multiple perspectives depending on the research objective. This can be predicting election results, a contest winner, or predicting peoples' reaction to a government decision through protest. The main perspectives to analyze the social networks are (1) content analysis, where the textual content of each post is analyzed using natural language processing to identify the topic or the sentiment of the authors. (2) Network structure analysis, where the relation between the users are described in a tree structure for the follower-followee patterns, or in a graph structure for friendship and interaction patterns. These patterns can be used to know the political preference of people prior to elections. (3) Behavioural analysis of each user including sentiment, response, likes, retweets, location, to identify responses toward specific events. This might be useful to identify users with terrorist intentions. In this section, we will focus on textual content-based models, where text analysis and understanding can be achieved using keywords, topic modelling or sentiment analysis. 
\subsection{Keyword-based approaches}
Keyword-based approaches focus on sequence analysis of the time series for each keyword. They also consider different forms for each keyword, including n-gram, skip-gram, and word-pairs \cite{fernandez2014gplsi}. The keyword-based approaches use the concept of the distributional semantics to group semantically-related words as synonyms to be used as a single feature \cite{landauer2006latent}. In this approach, keywords are usually associated with events by correlation, entropy or distance metrics. Also, Hossny \emph{et al.} proposed using SVD with K-Means to strengthen keyword signals, by grouping words having similar temporal patterns, then mapping them into one central word that has minimum distance to the other members of the cluster \cite{Hossny2018SVD}.

Sayyadi \emph{et al.} used co-occurring keywords in documents such as news articles to build a network of keywords. This network is used as a graph to feed a community detection algorithm in order to identify and classify events  \cite{sayyadi2009event}.  Takeshi \emph{et al.} created a probabilistic spatio-temporal model to identify natural disasters events such as earthquakes and typhoons using multiple tweet-based features such as words counts per tweet, event-related keywords, and tweet context. They considered each Twitter user as a social sensor and applied both of the Kalman filter and particle filter for location estimation. This model could detect 96\% of Japanese earthquakes  \cite{sakaki2010earthquake}. Zhou \emph{et al.} developed a named entity recognition model to find location names within tweets and use them as keyword-features for event detection, then estimated the impact of the detected events qualitatively   \cite{zhou2016real}. 

Weng \emph{et al.}  introduced ``Event Detection by Clustering of Wavelet-based Signals'' (EDCow). This model used wavelets to analyze the frequency of word signals, then calculated the autocorrelations of each word signal in order to filter outlier words. The remaining words were clustered using a modularity-based graph partitioning technique to form events \cite{weng2011event}. Ning \emph{et al.} proposed a model to identify evidence-based precursors and forecasts of future events.  They used as a set of news articles to develop a nested multiple instance learning model to predict events across multiple countries.  This model can identify the news articles that can be used as precursors for a  protest \cite{ning2016modeling}. 

\subsection{Topic modelling approaches}
Topic modelling approaches focus on clustering related words according to their meaning, and indexing them using some similarity metric such as cosine similarity or Euclidean distance. The most recognized techniques are (1) Latent Semantic Indexing (LSI), where the observation matrix is decomposed using singular value decomposition and the data are clustered using K-Means  \cite{landauer2006latent},(2) Latent Dirichlet Allocation (LDA), where the words are clustered using Gaussian mixture models (GMM) according to the likelihood of term co-occurrence within the same context \cite{sasaki2014online}, (3) Word2Vec, which uses a very large corpus to compute continuous vector representations, where we can apply standard vector operations to map one vector to another  \cite{mikolov2013distributed}.

Cheng \emph{et al.} suggested using space-time scan statistics to detect events by looking for clusters within data across both time and space, regardless of the content of each individual tweet  \cite{cheng2014event}. The clusters emerging during spatio-temporal relevant events are used as an indicator of a currently occurring event, as people tweet more often about event topics and news. Ritter \emph{et al.} proposed a framework that uses the calendar date, cause and event type to describe any event in a way similar to the way Twitter users mention the important events. This framework used temporal resolution, POS tagging, an event tagger, and named entity recognition. Once features are extracted,  the association between the combination of features and the events is measured in order to know what are the most important features and how significant the event will be   \cite{ritter2012open}.

Zhou \emph{et al.}  introduced a graphical model to capture the information in the social data including time, content, and location, calling it location-time constrained topic (LTT). They measure the similarity between the tweets using KL divergence to assess media content uncertainty. Then, they measure the similarity between users using a ``longest common subsequence'' (LCS) metric. They aggregate the two measurements by augmenting weights as a measure for message similarity. They used the similarity between streaming posts in a social network to detect social events \cite{zhou2014event}.

Ifrim \emph{et al.}  presented another approach for topic detection that combines aggressive pre-processing of data with hierarchical clustering of tweets. The framework analyzes different factors affecting the quality of topic modelling results  \cite{ifrim2014event}, along with real-time data streams of live tweets to produce topic streams in close to real-time rate.   

Xing \emph{et al.}  presented the mutually generative Latent Dirichlet Allocation model (MGE-LDA) that uses hashtags and topics, as they both are generated mutually by each other in tweets. This process models the relationship between topics and hashtags in tweets, and uses them both as features for event discovery \cite{xing2016hashtag}. Azzam \emph{et al.} used deep learning and cosine similarity to understand short text posts in communities of question answering\cite{Azzam2017_1,Azzam2017_2}. Also, Hossny et al. used inductive logic programming to understand short sentences from news for translation purposes \cite{hossny2009machine}     

\subsection{Sentiment analysis approaches}
The third approach is to identify sentiment through the context of the post, which is another application for distributional semantics requiring a huge amount of training data to build the required understanding of the context. Sentiment analysis approaches focus on recognizing the feelings of the crowd and use the score of each feeling as a feature to calculate the probability of social events occurring. The sentiment can represent the emotion, attitude, or opinion of the user towards the subject of the post. One approach to identify sentiment is to find smiley faces such as emoticons and emojis within a tweet or a post. Another approach is to use a sentiment labelled dictionary such as SentiWordNet to assess the sentiment associated with each word.

Generally, sentiment analysis has not been used solely to predict civil unrest, especially as it still faces the challenges of sarcasm and understanding negation in ill-formed sentences. Meanwhile, it is used as an extra feature in combination with features from other approaches such as keywords and topic modelling.
Paul \emph{et al.} proposed a framework to predict the results of the presidential election in the United States in 2017. The proposed framework applied topic modelling to identify related topics in news, then used the topics as seeds for Word2Vec and LSTM to generate a set of enriched keywords. The generated keywords will be used to classify politics-related tweets, which are used to evaluate the sentiment towards each candidate. The sentiment score trend is used to predict the winning candidate \cite{paul2017compass}.

\begin{figure*}
\centering
\includegraphics[width=0.7\linewidth,height=3.6cm]{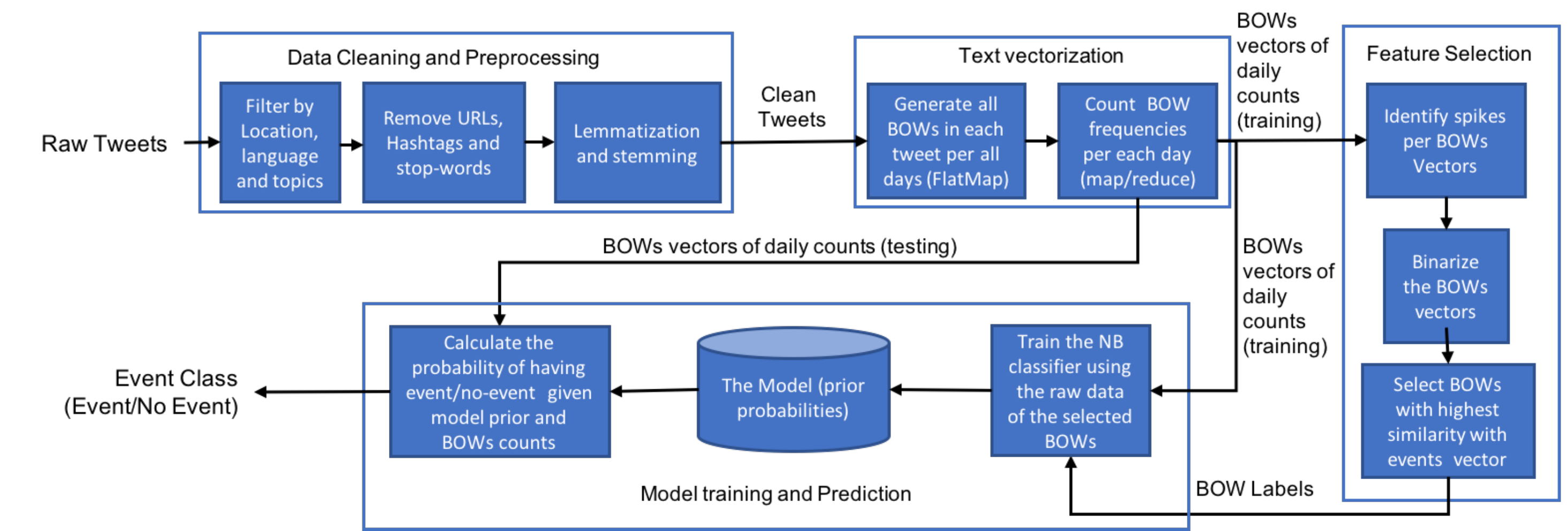}
\caption{The detailed pipeline for data processing, count vectorization, word-pair selection, training and prediction.}
\label{fig:detailedpipeline}
\end{figure*}

\section{Feature Selection Methods}

Keywords can be selected as features as a single term or a word-pair or a skip-grams, which can be used for classification using multiple methods such as mutual information, TF-IDF, $\chi^2$, or traditional statistical methods such as ANOVA or correlation. Our problem faces two challenges: the first is the huge number of word-pairs extracted from all tweets for the whole time frame concurrently, which make some techniques such as TF-IDF and $\chi^2$ computationally unfeasible as they require the technique to be distributable on parallel processors on a cluster. The second challenge is the temporal nature of the data which require some techniques that can capture the distributional semantics of terms along with the ground truth vector. In this section, we describe briefly a set of data association methods used to find the best word-pairs to identify the event days.

Pearson correlation measures the linear dependency of the response variable on the independent variable with the maximum dependency of 1 and no dependency of zero. This technique needs to satisfy multiple assumptions to assess the dependency properly. These assumptions require the signals of the variables to be normally distributed, homoskedastic, stationary and have no outliers \cite{benesty2009pearson,havlicek1976robustness}. In social network and human-authored tweets, we cannot guarantee that the word-pairs signals throughout the timeframe will satisfy the required assumptions. Another drawback for Pearson correlation is that zero score does not necessarily imply no correlation, while no correlation implies zero score. 

Spearman is a rank-based metric that evaluates the linear association between the rank variables for each of the independent and the response variables. It simply evaluates the linear correlation between the ranked variables of the original variables. Spearman correlation assumes the monotonicity of the variables but it relaxes the Pearson correlation requirements of the signal to be normal, homoskedastic and stationary. Although the text signals in the social network posts do not satisfy the monotonicity assumption, Spearman correlation can select some word-pairs to be used as predictive features for classification. Spearman correlation has the same drawback of Pearson correlation that zero score does not necessarily imply no correlation while no correlation implies zero score. 

Distance correlation is introduced by Szekely et al . (2007) to measure the nonlinear association between two variables \cite{szekely2007measuring}. Distance correlation measures the statistical distance between probability distributions by dividing the Brownian covariance (distance covariance) between X and Y by the product of the distance standard deviations\cite{szekely2009brownian,ayache2000covariance}.

TF-IDF is the short of \emph{ term frequency-inverse document frequency} technique that is used for word selection for classification problems. The concept of this technique is to give the words that occur frequently within a specific class high weight as a feature and to penalize the words that occur frequently among multiple classes. for example; the term ``Shakespeare'' is considered a useful feature to classify English literature documents as it occurs frequently in English literature and rarely occurs in any other kind of documents. Meanwhile, the term ``act'' will occur frequently in English literature, but it also occurs frequently in the other types of document, so this term will be weighted for its frequent appearance and it will be penalized for its publicity among the classes by what we call inverse-document-frequency\cite{ramos2003using}.

Mutual information is a metric for the amount of information one variable can tell the other one. MI evaluates how similar are the joint distributions of the two variables with the product of the marginal distributions of each individual variable, which makes MI more general than correlation as it is not limited by the real cardinal values, it can also be applied to binary, ordinal and nominal values\cite{fraser1986independent}. As mutual information uses the similarity of the distribution, it is not concerned with pairing the individual observations of X and Y as much as it cares about the whole statistical distribution of X and Y. This makes MI very useful for clustering purposes rather than classification purposes\cite{viola1997alignment}.

Cosine similarity metric calculates the cosine of the angle between two vectors. The cosine metric evaluates the direction similarity of the vectors rather the magnitude similarity. The cosine similarity score equals to 1 if the two vectors have the angle of zero between the directions of two vectors, and the score is set to zero when the two vectors are perpendicular\cite{crandall2008feedback}. if the two vectors are oriented to opposite directions, the similarity score is -1. Cosine similarity metric is usually used in the positive space, which makes the scores limited within the interval of [0,1]. 

Jaccard index or coefficient is a metric to evaluate the similarity of two sets by comparing their members to identify the common elements versus the distinct ones. The main advantage of Jaccard similarity is it ignores the default value or the null assumption in the two vectors and it only considers the non-default correct matches compared to the mismatches. This consideration makes the metric immune to the data imbalance. Jaccard index is similar to cosine-similarity as it retains the sparsity property and it also allows the discrimination of the collinear vectors.

\section{Spike Matching Method:}

The proposed model extracts the word-pairs having a high association with event days according to the distributional semantic hypothesis and use them for training the model that will be used later for the binary classification task\cite{mandera2017explaining} as illustrated in figure \ref{fig:detailedpipeline}. 
The first step is the data preparation where we load all the tweets for each day, then we exclude the tweets having URLs or unrelated topics, then we clean each tweet by removing the hashtags, non-Latin script and stopping words. Then we lemmatize and stem each word in each tweet using Lancaster stemmer. Finally, we extract the word-pairs in each tweet. The word-pair is the list of n words co-occurring together within the same tweet.

The second step is to count the frequency of each word-pair per each day, which are used as features to classify the day as either event or no-event day. The formulation is a matrix with rows as word-pairs and columns as days and values are daily counts of each word-pair. The third step is to binarize the event count vector (ground truth) as well as the vector of each word-pair. Binarizing the event vector is done by checking if the count of events in each day is larger than zero. The binarization of the word-pair count vectors is done by applying a temporal filter to the time series in order to identify the spikes as explained in equation \ref{eq:spike}, where the days with spikes are set to ones and days without spike are set to zeros \cite{franke2010online,sharafi2013information}. 

\begin{equation}
    f(x) = 
    \begin{cases}
    1, & \mbox{if } \mbox{ x(t)-x(t-1) $<$ threshold and} \\
    ~~  & \mbox{~~~~} \mbox{x(t)-x(t+1) $>$ threshold} \\
    0, &  \mbox{Otherwise} \\
    \end{cases}
\label{eq:spike}
\end{equation}

Where x is the count of the word-pair, $t$ is the time variable, $dt$ is the time difference, the threshold is the minimum height of the spike. Afterwards, we compare the binary vector for each word-pair with the ground truth binary vector using the Jaccard similarity index as stated in equation \ref{eq:jaccard} \cite{niwattanakul2013using,bank2008calculating}. The word-pairs are then sorted descendingly according to the similarity score. The word-pairs with the highest scores are used as a feature for training the model in the fourth step. 

\begin{equation}
\begin{split}
J(word-pair,GT) = \dfrac{WP \cap GT}{WP \cup GT} =  \dfrac{\sum_{i} \min({WP_i,GT_i}) }{\sum_{i} \max({WP_i,GT_i}) }
\end{split}
\label{eq:jaccard}
\end{equation}

where WP is the word pair vector, GT is the ground truth vector 

\section{Training and Prediction}
Once we identify the best word-pairs to be used as features for classification, we split the time series vector of each word-pair into a training vector and a testing vector. then we use the list of the training vectors of the selected word-pairs to train the model as explained in subsection \ref{subsec:training} and use the list of testing vectors for the same word-pairs to classify any day to event/nonevent day \ref{subsec:predict}.

\subsection{Training the model:}
\label{subsec:training}

The third step is to train the model using the set of features generated in the first step. We selected the Naive Bayes classifier to be our classification technique for the following reasons: (1) the high bias of the NB classifier reduces the possibility of over-fitting, and our problem has a high probability of over-fitting due to the high number of features and the low number of observations, (2) the response variable is binary, so we do not need to regress the variable real value as much as we need to know the event-class, and (3) The counts of the word-pairs as independent variables are limited between 0 and 100 occurrences per each day, which make the probabilistic approaches more effective than distance based approaches.  

The training process aims to calculate three priori probabilities to be used later in calculating the posterior probabilities: (1) the probability of each word-pair count in a specific day given the status of the day as ``event'' or ``non-event''. (2) the priori conditional probability of each word-pair given event status $P(word-pair|Event)$. (3) the probability of each event class as well as the probability of each word-pair as stated in equations \ref{eq:prob_word-pair} and \ref{eq:prob_word-pair}.  
\begin{equation}
P(Event_c) = \dfrac{Count(Event_c)} {\sum_{c\in\{0,1\}} Count(Event_c)}
\label{eq:prob_events}
\end{equation}
\begin{equation}
P(WP|Event_c) = \dfrac{P(WP \cap Event_c)} {P(Event_c)}
\label{eq:prob_word-pair}
\end{equation}
\noindent
where $WP$ is the word-pair, ${Event}_c$ is any class for event occurrence and word-pair is the vector of counts for the word-pairs extracted from tweets
\subsection{Predicting Civil Unrest}
\label{subsec:predict}
Once the priori probabilities are calculated using the training data, we use them to calculate the posterior probability of both classes of event-days and non-event-days given the values of the word-pairs using the equation \ref{eq:posteriori}. 
\begin{equation}
\begin{split}
    P(Event_c|WP_1,WP_2, \dots, WP_n) \\
    = \dfrac{P(WP_1,...,WP_n).P(Event)}{P(WP_1,\dots,WP_n)} \\
    = \dfrac{1}{Z} P(Event) \prod_{i=1}^{n} P(WP_i|Event)
\end{split}
\label{eq:posteriori}
\end{equation}
\noindent
where 
$WP$ is the word-pair, $Z= P(WP1,\dots)$ $=P(WP_1).P(WP_2). \dots$
As the word-pairs are assumed to be independent and previously known from the training step.

\begin{table*}
\centering
\caption{Event detection results for 10 randomized folds using the metrics of accuracy, precision, recall, F1, Area under ROC curve and area under PR curve}
\label{tbl:cormethods}
\begin{tabular}{|l|l|l|l|l|l|l|}
\hline
method              & AUC ROC      & F1s           & AUPR                      & Precision & Recall      & Accuracy \\ \hline
Pearson correlation & 0.759        & 0.517         & 0.457                     & 0.658     & 0.431       & 0.758         \\ \hline
distance correlation            & 0.831        & 0.677         & 0.564                     & 0.687     & 0.669       & 0.807         \\ \hline
Mutual\_Information        & 0.552        & 0.426         & 0.321                     & 0.328     & 0.613       & 0.504         \\ \hline
Jaccard similarity& 0.894        & 0.794         & 0.676                     & 0.727     & 0.878       & 0.862         \\ \hline
Cosine similarity  & 0.651        & 0.397         & 0.358                     & 0.431     & 0.405       & 0.653         \\ \hline
Spike matching     & 0.913        & 0.793         & 0.689                     & 0.770     & 0.821       & 0.873         \\ \hline

\end{tabular}
\end{table*}

\begin{table*}[]
\centering
\caption{A comparison of classification AUCs using word-pairs extracted by different feature selection methods}
\label{tbl:clf_aucs}
\begin{tabular}{|l|l|l|l|l|l|l|l|l|l|}
\hline
method              & NB    & SVM   & MLP   & LDA   & GPC   & LR    & RF    & DT    & KNN   \\ \hline
Pearson Correlation  & 0.759 & 0.555 & 0.778 & 0.563 & 0.598 & 0.784 & 0.740 & 0.720 & 0.614 \\ \hline
Distance correlation            & 0.831 & 0.589 & 0.874 & 0.713 & 0.667 & 0.875 & 0.866 & 0.814 & 0.637 \\ \hline
Mutual Information        & 0.558 & 0.554 & 0.509 & 0.525 & 0.502 & 0.539 & 0.559 & 0.531 & 0.511 \\ \hline
Jaccard Similarity & 0.894 & 0.659 & 0.951 & 0.795 & 0.598 & 0.955 & 0.934 & 0.921 & 0.691 \\ \hline
Cosine Similarity  & 0.651 & 0.561 & 0.591 & 0.527 & 0.497 & 0.670 & 0.528 & 0.600 & 0.523 \\ \hline
Spike Matching     & 0.913 & 0.517 & 0.963 & 0.759 & 0.891 & 0.966 & 0.965 & 0.929 & 0.657 \\ \hline
\end{tabular}
\end{table*}

\section{Experiments and Results}

The experiments are designed to detect civil unrest events in Melbourne on any specific day. In this experiment, we used all the tweets posted from Melbourne within a time frame of 640 days between December 2015 and September 2017. This time frame will be split into 500 days for model training and 140 days for model testing on multiple folds. The tweet location is specified using (1) longitude and latitude meta-tag, (2) tweet location meta-tag, (3) the profile location meta-tag, and (4) The time zone meta-tag. The total number of tweets exceeded 4 million tweets daily. Firstly, we cleaned the data from noisy signals, performed stemming and lemmatization then extracted the word-pairs from each tweet and count each word-pair per each day. Example 1 illustrates how each tweet is cleaned, prepared and vectorized before being used for training the model. The steps are explained below:
\begin{itemize}
\item The data is first cleaned by eliminating the tweets of any language other than English, exclude the tweets having URLs, remove the hash-tags, non-Latin alphabets, punctuation, HTML Tags, and remove the stopping words listed by NLTK \cite{Loper:2002NLTK}.
\item Extract the word-pairs from each tweet by generating a list of every two co-occurring words. The number of word-pairs extracted from a tweet with size $m$ equals $m*(m-1)$ . Each tweet consists of average 12 words, which construct 132 word-pairs per tweet on average. The average count of daily different word-pairs exceeds 10 million after excluding repeated word-pairs and the word-pairs with single appearance.
\item Tإhe  words in each word-pair are lemmatized using NLTK lemmatizer in order to avoid the morphological effects to the word shape (e.g. bought $\rightarrow$ buy) 
\item The words in each word-pair are then stemmed by Lancaster stemmer in order to return related words to their dictionary roots (E.g., Turkish $\rightarrow$ Turk)
\item The word-pairs are then counted per day for all the tweets from Melbourne in order to construct the term frequency vectors.
\end{itemize}

As explained in example 1, each word-pair will be transformed from a vector of integer values into a vector of binary values and denoted as ${B\widetilde{O}W}$. ${B\widetilde{O}W}$ will be used to calculate the Jaccard similarity index of the binary vector with the events binary vector. Each word-pair will have a similarity score according to the number of word-pair spikes matching the event days. This method uses the concept of distributional semantic, where the co-occurring signals are likely to be semantically associated \cite{mandera2017explaining}. 

\noindent
\fbox{\begin{minipage}[t]{.48\textwidth}%
\textbf{Example 1:}\\
\textbf{Original Tweet:} \\ 
Protesters may be unmasked in wake of Coburg clash https://t.co/djjVIfzO3e (News) \#melbourne \#victoria\\

\textbf{Cleaned Tweet:} \\
protest unmask wake coburg clash news\\

\textbf{List of two-words-word-pairs:} [`protest', `unmask'], [`protest', `wake'], [`protest', `Coburg'], \dots, [`unmask', `wake'], [`unmask', `coburg'],\dots , [`clash', `news'] 
\\

[`protest', `unmask'] training : $[x_{1,1},x_{1,2},x_{1,3},\ \ldots\ ,x_{1,641}]$ 
[`protest', `unmask'] testing : $[x_{2,1},x_{2,2},\ \ldots\ ,x_{1,641}]$ 
\\

Assuming a time frame of 20 days \\
word-pair:  [2,3,3,4,5,3,2,3,8,3,3,1,3,9,3,1,2,4,5,1] \\
Spikes (${B\widetilde{O}W}$): [0,0,0,0,1,0,0,0,1,0,0,0,0,1,0,0,0,0,1,0]\\
Events($GT$): [0,0,0,0,1,0,0,0,0,0,0,0,0,1,0,0,1,0,1,0]\\

$J({B\widetilde{O}W},GT) = \dfrac{\sum_{i} \min({{B\widetilde{O}W}_i,GT_i} )}{\sum_{i} \max({{B\widetilde{O}W}_i,GT_i}) } = \dfrac{3}{5}$

\end{minipage}}
\\
\\

\begin{figure*}[h!]
\centering
\begin{subfigure}{.24\textwidth}
  \centering
  \includegraphics[width=1\linewidth]{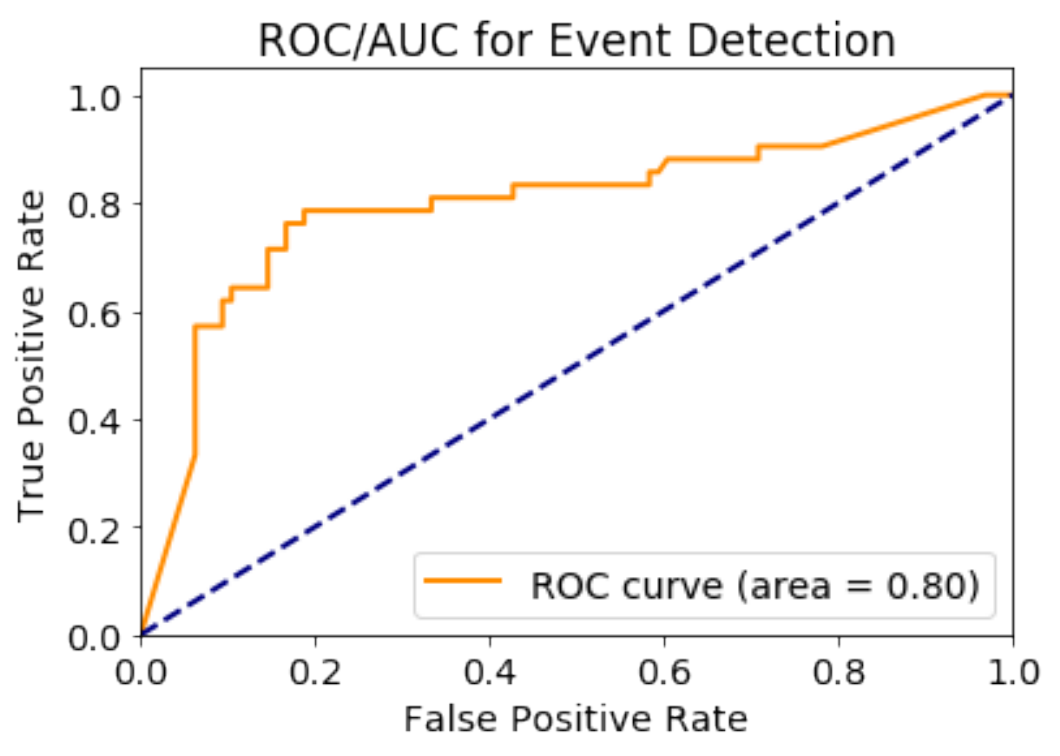}
\caption{Pearson Correlation ROC }
\label{fig:ROC}
\end{subfigure}%
\begin{subfigure}{.24\textwidth}
  \centering
  \includegraphics[width=1\linewidth]{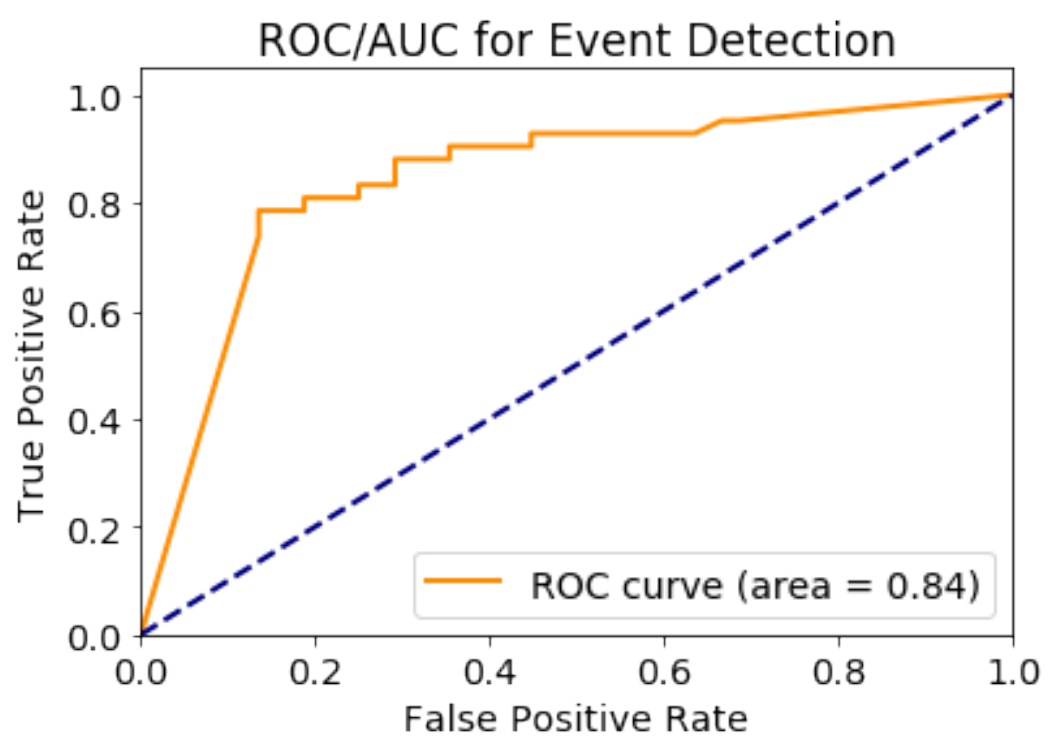}
\caption{Distance correlation ROC}
\label{fig:ROC}
\end{subfigure}%
\begin{subfigure}{.24\textwidth}
  \centering
  \includegraphics[width=1\linewidth]{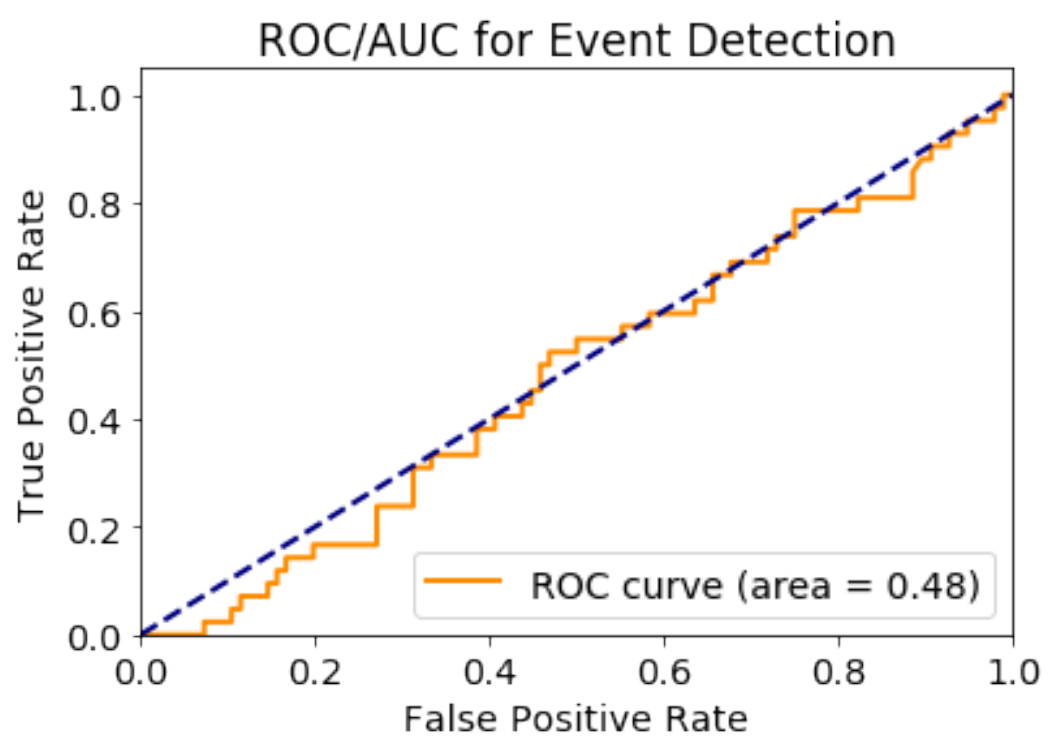}
\caption{Mutual-Information Roc}
\label{fig:ROC}
\end{subfigure}%
\hfill
\begin{subfigure}{.24\textwidth}
  \centering
  \includegraphics[width=1\linewidth]{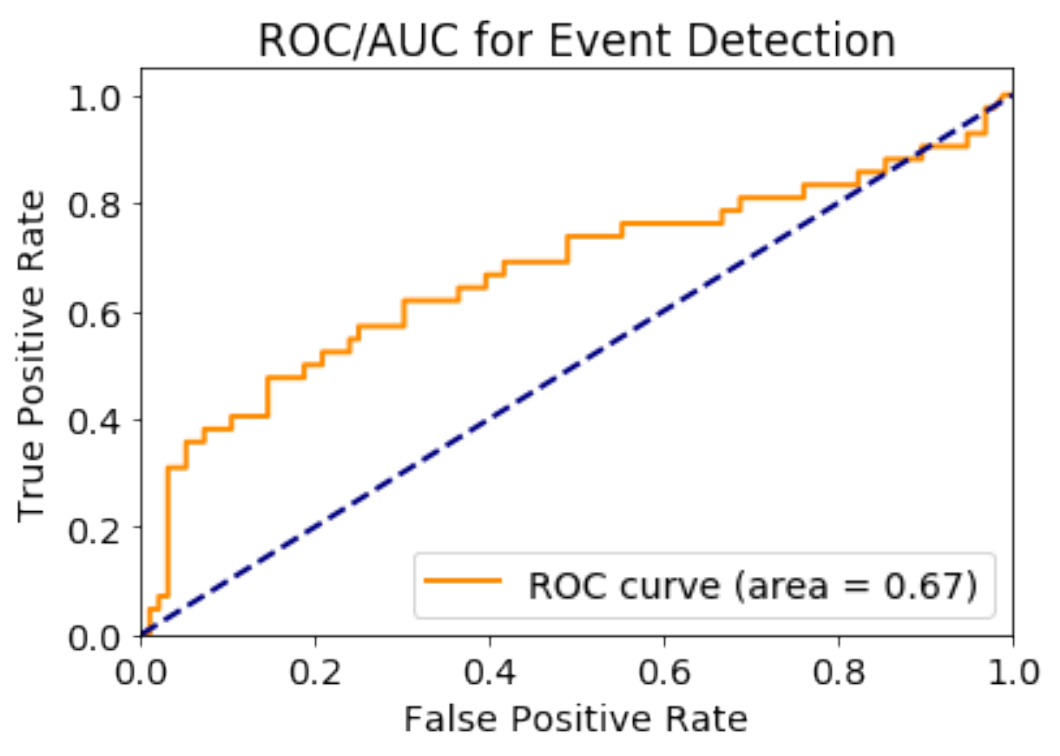}
\caption{Cosine Similarity ROC}
\label{fig:ROC}
\end{subfigure}%
\begin{subfigure}{.24\textwidth}
  \centering
  \includegraphics[width=1\linewidth]{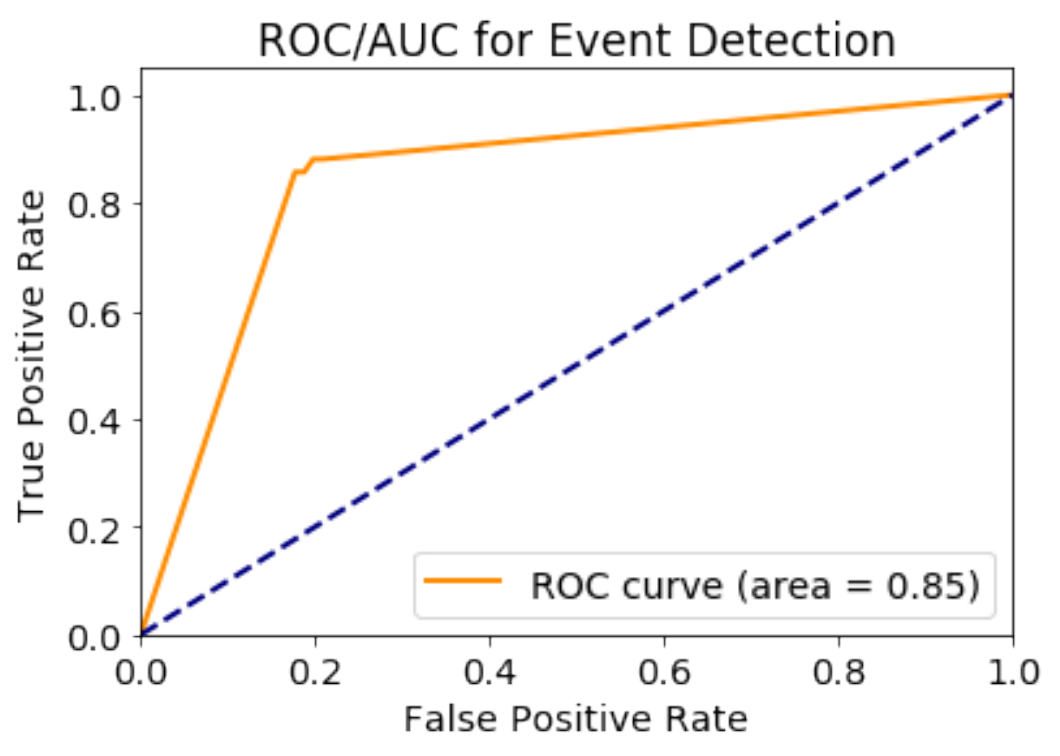}
\caption{Jaccard Similarity ROC}
\label{fig:ROC}
\end{subfigure}%
\begin{subfigure}{.24\textwidth}
  \centering
  \includegraphics[width=1\linewidth]{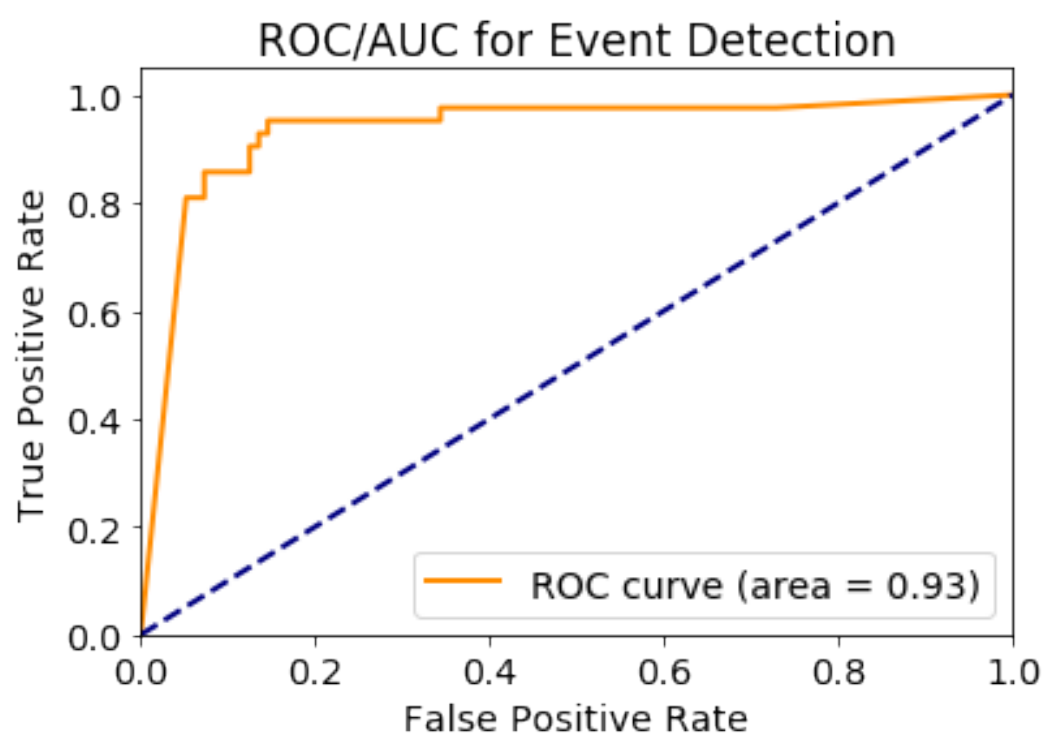}
\caption{Spike Matching ROC}
\label{fig:ROC}
\end{subfigure}%
\caption{ROC curves for Naive Bayes classification using features extracted by different selection methods }
\label{fig:ROCs}

\end{figure*}

\begin{table*}[]
\centering
\caption{The average results of event detection in multiple cities using multiple metrics after cross validating the results on 10 folds}
\label{tbl:countries}
\begin{tabular}{|p{1.4cm}|p{1.4cm}|p{1.4cm}|p{1.1cm}|p{0.9cm}|p{1.5cm}|p{1.6cm}|}
\hline
City      & Accuracy & Precision & Recall & F1    & AUC PR & AUC ROC \\ \hline
Melbourne  & 0.873    & 0.770     & 0.820  & 0.793 & 0.688   & 0.913    \\ \hline
Sydney    & 0.860    & 0.707     & 0.770  & 0.719 & 0.640   & 0.897    \\ \hline
Brisbane  & 0.855    & 0.514     & 0.612  & 0.528 & 0.449   & 0.791    \\ \hline
Perth     & 0.903    & 0.605     & 0.797  & 0.686 & 0.574   & 0.886    \\ \hline
Jakarta   & 0.762    & 0.816     & 0.706  & 0.705 & 0.735   & 0.860    \\ \hline
\end{tabular}
\end{table*}

Once we selected the most informative word-pairs as features, we will use the raw values to train the Naive Bayes classifier. The classifier is trained using 500 days selected randomly along the whole timeframe, then it is used to predict the other 140 days. To ensure the robustness of our experiment, We applied 10-folds cross-validation, where we performed the same experiment 10 times using 10 different folds of randomly selected training and testing data. The prediction achieved an average area under the ROC curve of 90\%, which statistically significant and achieved F-score of 91\%, which is immune to data imbalance as listed in table \ref{tbl:cormethods}. Figure \ref{fig:ROCs} shows the ROC curves for the results of a single fold of Naive Bayes classification that uses the features extracted by each selection methods. The classification results of the proposed method outperformed the benchmarks and state of the art developed by  Cui et al. (2017), Nguyen et al. (2017), Willer et al. (2016),  and Adedoyin-Olowe et al. (2016) as illustrated in the table \ref{tbl:benchmarks} \cite{weng2011event, cordeiro2012twitter,WEILER2016207,CUI201753,ADEDOYINOLOWE2016351,NGUYEN2017137}.

\begin{table*}[]
\centering
\caption{The classification scores compared to benchmarks }
\label{tbl:benchmarks}
\begin{tabular}{|p{3.5cm}|p{0.4cm}|p{7.4cm}|p{0.9cm}|p{0.9cm}|}
\hline
Author                       & Ref                                          & Method                                                                                                         & F-score & Accuracy \\ \hline
Cui et al. (2017)            & \cite{CUI201753}            & used a graph based key-phrase extraction algorithm  (TextRank) to extract the key-phrases matching the events. &         & 0.65     \\ \hline
Weng et al. (2011))          & \cite{weng2011event}        & EDCoW: Clustering of Wavelet-based Signals followed by discrete wavelet analysis for each term                 & 0.43    &          \\ \hline
Crodeiro et al. (2012)       & \cite{cordeiro2012twitter}  & WATIS: Wavelet Analysis Topic Inference Summarization                                                          & 0.43    &          \\ \hline
Adedoyin-Olowe et al. (2016) & \cite{ADEDOYINOLOWE2016351} & Transaction-based Rule Mining to extract worthy hashtag keywords                                               & 0.77    &          \\ \hline
Proposed method              &                                              & Keyword volume and spike matching approach                           & 0.79    & 0.87     \\ \hline
\end{tabular}
\end{table*}

The same experiment has been applied to Sydney, Brisbane and Perth in Australia on a time frame of 640 days with 500 days training data and 140 days testing data and the results were similar to Melbourne results with average AUC of 0.91 and average F-Score of 0.79. To ensure that the proposed method is language independent, we used the same method to classify civil unrest days in Jakarta using the Indonesian language, the classification scores were lower than the average scores for English language by 0.05 taking into consideration that we did not apply any NLP pre-processing to the Indonesian tweets such as stemming and lemmatization. 

To verify the robustness of this feature selection method, we tested the selected features using multiple classifiers such as KNN, SVM, naive Bayes and decision trees. The results emphasized that the word-pairs selected using the spike-matching method achieve better AUC scores than the other correlation methods as listed in table \ref{tbl:clf_aucs}


\section{Conclusions}
In this paper, we proposed a framework to detect civil unrest events by tracking each word-pair volume in twitter. The main challenge with this model is to identify the word-pairs that are highly associated with the events with predictive power. We used temporal filtering to detect the spike within the time series vector and used Jaccard similarity to calculate the scores of each word-pair according to its similarity with the binary vector of event days. These scores are used to rank the word-pairs as features for prediction.

Once the word-pairs are identified, we trained a Naive Bayes classifier to identify any day in a specific region to be an event or non-event days. We performed the experiment on both Melbourne and Sydney regions in Australia, and we achieved a classification accuracy of 87\% with the precision of 77\%, Recall of 82 \%, area under the ROC curve of 91\% and F-Score of 79\%. The results are all achieved after 10-folds randomized cross-validation as listed in table \ref{tbl:countries}. 

The main contributions of this paper are (1) to overcome twitter challenges of acronyms, short text,  ambiguity and synonyms, (2) to identify the set of word-pairs to be used as features for live event detection, (3) to build an end-to-end framework that can detect the events lively according to the word counts. This work can be applied to similar problems, where specific tweets can be associated with life events such as disease outbreak or stock market fluctuation. This work can be extended to predict future events with one day in advance, where we will use the same method for feature selection in addition to to time series analysis of the historical patterns of the word-pairs.   

\section{Acknowledgments}
This research was fully supported by the School of Mathematical Sciences at the University of Adelaide. All the data, computation and technical framework were supported by Data-To-Decision-Collaborative-Research-Center (D2DCRC).
%
\bibliographystyle{abbrv}
\bibliography{sigproc}  
%
%
\end{document}